# CAGE OCCUPANCIES OF METHANE HYDRATES: RESULTS FROM SYNCHROTRON X-RAY DIFFRACTION AND RAMAN SPECTROSCOPY

**Junfeng QIN, Christiane D. HARTMANN and Werner F. KUHS**[*]
GZG Abt. Kristallographie, Georg-August-Universität Göttingen, Göttingen 37077, Germany

**ABSTRACT**

An accurate knowledge of cage occupancy of methane is central for understanding the physical-chemical properties of gas hydrates, the actual inventory of natural gas in hydrate deposits and the description of gas exchange processes. Here we report the absolute cage occupancies, the cage occupancy ratios and hydration numbers of the synthetic $CH_4$-$H_2O$ and $CH_4$-$D_2O$ hydrates formed from the ice-gas system under different pressures and temperatures. The results were obtained from Rietveld refinement using high-resolution synchrotron X-ray powder diffraction patterns and from Raman spectroscopic measurements. The small-cage occupancies of methane in the deuterated hydrates are found to be slightly higher than in the hydrogenated form, likely due to their different lattice constants. The $CH_4$ occupancy in the small cages agrees fairly well with the predictions of CSMGem at the formation pressure of 3.5 MPa, but with the increasing formation pressure the disagreement grows up to 11 percent. While some deficiency of the prediction model cannot be excluded, the observed discrepancy may well be due to experimental difficulties of reaching true equilibrium at higher pressures. The experimentally-determined large-to-small cage occupancy ratios of the synthetic and natural $CH_4$ hydrates formed from the water-gas system are consistently higher than the results of CSMGem calculations. Possible reasons for these discrepancies will be discussed.

**INTRODUCTION**

Natural gas hydrates (NGHs) have been attracting an increasing attention due to the potential as an energy resources, their storage capacity for the greenhouse gas $CO_2$, as well as their relevance to submarine geohazards and global climate change, e.g. [1-3]. Gas hydrates are also the origin of costly blockages of gas and oil pipelines operated at the elevated pressures [4]. These energy-related, environmental and engineering concerns require a detailed understanding of structure, phase equilibrium and composition of gas hydrates.

Gas hydrates are non-stoichiometric crystalline materials in which small molecules are enclathrated into hydrogen-bonded water cages via van der Waals-type guest-host interactions. Methane hydrates, the most abundant naturally-occurring gas hydrates, are usually found to form a cubic structure I (sI) [5]. The unit cell of sI hydrate is comprised of two small cages ($5^{12}$, SCs) and six large ones ($5^{12}6^2$, LCs) with a total of 46 water molecules. Methane molecule occupies both cages, but the cage occupancies are non-stoichiometric with some vacancies. The cage filling depends on the fugacity of the guest species in the gas phase during formation/equilibration as well as on the molecular interactions between the guest and the host lattice. Therefore, the cage occupancy $\theta$ can give insight into the fundamentally physical-chemical interactions in gas hydrates. The general chemical formula of methane hydrates is $CH_4 \cdot n H_2O$, where $n$ is the hydration number referring to molar water molecules per methane molecule in methane hydrates. For the limiting fully occupied sI $CH_4$ hydrate, $n$ is equal to 5.75. The hydration number can be experimentally derived in three ways: the experimental determination of cage-occupancies [6-8], a thermodynamic analysis of phase equilibria data [9,10] or a direct measurement of the water-to-gas ratio after $CH_4$ hydrate dissociation [4,11-13] or gas uptake during hydrate formation. Methane hydrate hydration numbers were reported over a wide range from 5.77 to 7.40 [4,6-13]. The direct measurement method suffers from some problems, such as the occlusion of the unreacted ice or water in the hydrate crystals during formation [13], and the new formed ice arising from hydrate dissociation and the trapped interstitial sea water during core recovery [14]. Methane hydrate stoichiometry determined by an experimental determination of the cage occupancies is not directly affected by the presence of ice. The hydration number of sI $CH_4$ hydrates can be calculated by,

$$n = \frac{23}{3\theta_{LC} + \theta_{SC}} \quad (1)$$

[*] Corresponding author: Phone: +49 551 393891 Fax +49 551 399521 E-mail: wkuhs1@gwdg.de

The accurate determination of the cage occupancy and hydration number allows for a better prediction of the actual inventory of natural gas in NGH reservoirs [1,15].

A large number of experimental and theoretical studies have addressed various aspects of cage occupancy and phase equilibria of gas hydrates. X-ray diffraction, nuclear magnetic resonance (NMR) and Raman spectroscopy are the most effective and widely used techniques for characterizing guest distributions in hydrate cavities. Diffraction is the only method to provide the accurate absolute cage occupancies of gas hydrates without further assumption or calibration. Single crystal X-ray diffraction is a powerful tool [16], yet sample preparation is more complex than for powder samples due to the generally rather small crystallite size leaving some scope for powder diffraction [17]. Only a few studies using single-crystal diffraction were reported for synthetic and natural gas hydrates [17-19]. X-ray powder diffraction is generally more suitable for gas hydrate studies, but also bound with problems resulting from the extensive disorder of both host lattice and guest positions. Particularly severe are parameter correlations among guest positional coordinates, guest occupancy and the thermal displacement parameters of the guests. In order to disentangle these correlations usually one needs to fix some of these parameters which in turn will cast doubts on the remaining freely refined values. Yet, most problems with these parameter correlations can be overcome by using very high quality synchrotron powder diffraction data extending to large scattering angles [20].

Spectroscopic techniques, like Raman and solid-state NMR, are powerful tools to investigate the small-to-large cage occupancy ratio of $CH_4$ in gas hydrates due to the good separation of $CH_4$-contributions in the LCs and SCs. Raman spectroscopy features high spatial resolution, whereas the compositional information given by Raman needs to calibrate the Raman scattering cross section for each species [6-8,21-33]; NMR is thought to provide more quantitative information [29,32,34-39], yet is often less accessible. The agreement on the cage occupancy ratio of $CH_4$ in sI $CH_4$ hydrates [29,34,35], cross calibrated by NMR and Raman, indicates that Raman scattering is a reliable tool, at least for sI $CH_4$ hydrates. In general, the absolute cage occupancies can be obtained by the integrated peak intensities of $CH_4$ in the LCs and SCs with thermodynamic constraints deduced from the van der Waals and Platteeuw (vdWP) model [40]. Recently, Raman scattering has been proposed as an independent tool to quantify gas hydrates by calibrating the relative Raman cross sections of guests and host cavities in the hydrate phase [6].

Unfortunately, only a limited number of experimental results performed under various gas hydrate formation conditions are available and they are not always consistent with each other. A variety of thermodynamic models have been developed for predicting phase equilibria and composition of gas hydrates [41-50]; most of them are based on the statistical theory proposed by van der Waals and Platteeuw (vdWP) [40]. The main difference of these models is the way to deduce the guest-host interaction potential. Parrish and Prausnitz generalized the original vdWP model and applied it for the multicomponent gas hydrates using the Kihara spherical core potential [41]. Ballard and Sloan improved the prediction by considering the distortion of the hydrogen-bonded hydrate cavities changing as a function of guest molecules [43]. Incorporating this modified vdWP model along with performing a multiphase Gibbs energy minimization, the user-friendly program, CSMGem, was developed and has been widely used in industrial settings [44]. One of main disadvantages of the Kihara potential is that the parameters need to be regressed from the available experimental data of phase equilibrium and cage occupancy, which makes the quality of the results crucially dependent on the accuracy of the experimental data. Recent work has suggested that intermolecular potentials directly calculated by *ab initio* quantum mechanical methods improve the prediction of hydrate properties [45-50]; the predicted cage occupancies are quite sensitive to the chosen intermolecular potential, even for pure methane hydrates [51]. Molecular simulation techniques are also useful tools to probe phenomena at the molecular level in gas hydrates and to assess the validity of the underlying approximations built into the vdWP theory, e. g. [52-54], yet these methods require more computation time and capacity [5].

Although the original vdWP model and its subsequent modifications have predicted the dissociation pressure of methane hydrates with some success [46], the observed discrepancies in cage occupancies between experiments and models imply that our understanding still needs to be improved [55], especially for the likely overestimation of $CH_4$ in the SCs [46,48]. This disagreement arises from the assumptions built into the vdWP theory and/or the scarce and inconsistent experimental values of the cage occupancy of $CH_4$.

In this study, the hydrogenated and deuterated sI $CH_4$ hydrates formed under different isobaric-isothermal conditions are investigated. The absolute cage occupancy, cage occupancy ratio and hydration number of $CH_4$ hydrates were determined by the measurements of high resolution synchrotron X-ray diffraction as well as Raman spectroscopy on identical samples. Then, our results are compared with the predictions of thermodynamic models, and the available experimentally-determined cage occupancy ratios of synthetic and natural methane hydrates (NMHs). These comparisons may help to improve our understanding of methane hydrate formation from the ice/water-gas system and lead to a more accurate model to predict cage fillings of gas hydrates.

**EXPERIMENTAL SECTION**

Methane hydrates were prepared from hexagonal $H_2O$ or $D_2O$ ice and $CH_4$ gas (purity 99.995%) under isobaric and isothermal conditions. The spherical $H_2O$ or $D_2O$ (purity 99.9% deuterated) ice particles with a typical diameter of tens of μm were formed by spraying water into liq. $N_2$ [56]. $D_2O$ ice was produced in a glovebox under dry $N_2$ atmosphere to prevent the isotopic contamination from the atmospheric $H_2O$. Ice particles were filled into an Al-vial with an inner diameter of 6.7 mm or in a larger PFA-jar, which was inserted into a precooled custom-built pressure vessel where temperature was controlled by a circulating cryostat bath. After air and residual nitrogen were eliminated from the vessel by flushing with methane for several times, the sample cell was immediately pressurized to the designated pressure; the pressure was continuously monitored with a pressure transducer (Ashcroft) calibrated to a mechanical high-precision Heise gauge. During the reaction, pressure was manually maintained to the set point within 0.1-0.2 MPa. All formation reactions ran for 3 weeks. At the completion of the reaction, the pressure cell was rapidly quenched in liq. $N_2$ and pressure was concomitantly released. The recovered samples were ground and sieved under liq. $N_2$ for the following synchrotron powder diffraction and Raman measurements, and stored under liquid $N_2$.

The powder diffraction data were acquired on the high-resolution diffractometer ID31 at ESRF (Grenoble, France) equipped with a nine crystal multi-analyzer stage. The wavelength was determined to be 0.403027 Å by using a silicon standard powder. The sample holder, a small quartz glass capillary with an inner diameter of 1.0-1.7 mm, was mounted vertically to the synchrotron beam (Bragg-Brentano geometry; $2\theta$-range of at least 0-48°, in some cases 0-100°) and spun with 300 rpm to improve grain statistics at each measured step. In addition, measurements from different sample positions and within different $2\theta$-ranges were taken and merged together. To prevent sample from decomposing and to achieve a low-noise diffraction pattern, the samples were cooled by a coaxial $N_2$ stream set to a nominal temperature of 100 K. The actual temperature at the sample was ~ 135±10 K, depending on the location of the sample in the $N_2$ stream.

Raman measurements were performed using a LabRAM HR800 (Horiba Jobin Yvon) Raman spectrometer equipped with a Peltier-cooled CCD detector (DU 420A, Andor). The instrumental parameters were set to: 600 grooves/mm grating, Ar+ laser (Innova 90C, Coherent) emitting wavelength of 488 nm at the output power of 20.5 mW, 50× long-working distance objective (Olympus) and 100 μm confocal hole. The laser beam focused on the surface of hydrate samples within a diameter of around 1.1 μm and the Raman signals collected in backscattering geometry (180°). These configurations allow Raman spectra to be collected with a spectral resolution of 2.2 cm$^{-1}$. Spectrum was acquired in an averaged two accumulations of 30 s exposure time. Hydrate samples were placed in a cooling stage (Linkam THMS600) and measured at 113 ±0.1 K under the ambient pressure of liq. N$_2$. The Raman peaks were fitted in the region of interest as described in [6].

## STRUCTURAL ANALYSIS

All diffraction data were analyzed in full-pattern crystallographic structure refinements using the program GSAS [57]. Zero shift, lattice parameters and angle-dependent profile functions with Lorentzian and Gaussian components were refined, see Figure 1. The background could not be properly fitted by the implemented functions because of the highly irregular diffuse scattering by the glass capillaries. Hence background points were set manually and linearly interpolated in GSAS. Structural parameters could be determined with high precision because of a very high reflection to parameter ratio and the very high resolution in 2$\theta$ (nominal instrumental contribution to peak broadening was 0.003°). Isotropic atomic displacement factors ($U_{iso}$) and in some cases even anisotropic ones ($U_{aniso}$) of guest molecules and cage fillings could be refined simultaneously, which was possible due to the large useful 2$\theta$-range of 0-50° (sin $\theta/\lambda$: 1.05 Å$^{-1}$) with 1472 unique observed hydrate reflections.

The initial structural parameters for the framework atoms were taken from neutron diffraction results obtained for CH$_4$-D$_2$O hydrates synthesized at 6 MPa and 273 K [58]. Oxygen framework positions and their $U_{iso}$'s were refined first, and finally new framework hydrogen positions were calculated. The neutron-derived positions were taken and shortened to O-H/D-distances of 0.7 Å, as it is customary for X-ray data to account for the electron density maximum of H-bonded H atoms.

The guest models for the LCs and SCs were also taken from [58]. The C-atoms were set into the middle of the cages (C$_{SC}$ = 0 0 0 and C$_{LC}$ = 0.25 0.5 0), and the H-atoms were arranged in such a way – that they were on crystallographic positions with maximized multiplicity to simulate a surface of a sphere. Starting $U_{iso}$'s were taken from [58], whereupon $U_{iso}$'s of C-atoms were reduced by half and the $U_{iso}$'s of H-atoms by 10% accounting for the low temperature during the measurements. At first, cage occupancies were refined freely. In case the refined large cage occupancy exceeded slightly 100%, the large cage filling was set back to 100%. The isotropic displacement parameter was fixed for the C-atom in the SC to a value close to the displacement parameter of the framework oxygen atoms, but could be refined anisotropically for the C-atom in the LC. The improvement of the fit by refining the $U_{aniso}$'s of the C-atom in the LC was found to be statistically significant [59].

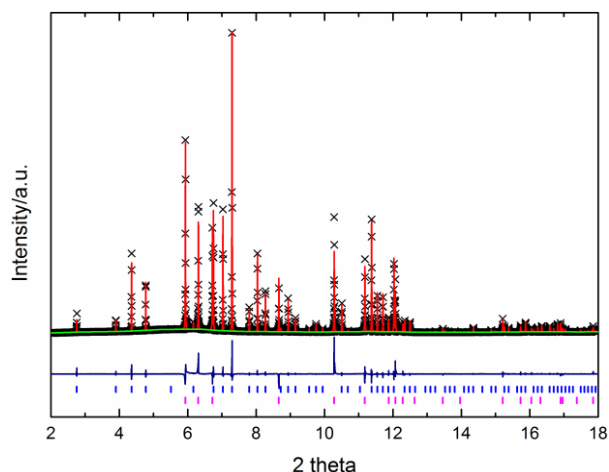

Figure 1 Rietveld plot of the synchrotron diffraction data of sI CH$_4$-H$_2$O hydrates synthesized at 6 MPa within the 2$\theta$ range of 2-18°. The top and bottom bars represent the Bragg peak positions of CH$_4$ hydrates and ice Ih, respectively. The bottom line corresponds to the difference between the observed and calculated patterns.

## RESULTS AND DISCUSSION
### Synchrotron X-ray diffraction

Structural parameters of framework and guest molecules can be accurately refined from the high-quality synchrotron diffraction data. Because the position and disorder of CH$_4$ molecules is well described by our model and by using anisotropic atomic displacement parameters for the C-atom in the LC, all small cage occupancies could be refined freely in most cases. Some of the LC occupancies refined to values larger than 100% and were subsequently fixed to 100%. The most important structural and compositional results of CH$_4$ hydrates determined by the analysis of synchrotron diffraction patterns are listed in Table 1. As it can be seen, the weight percentages of ice in the recovered $CH_4$ hydrates are very high. It means that the conversion of ice to $CH_4$ hydrates is not complete after 3 weeks of gas-ice reactions. The results show that the LCs are nearly full, while 11-15% of SCs are vacant. For the deuterated $CH_4$ hydrates, the small cage occupancy increases somewhat with the increasing formation pressure, yet the hydrogenated $CH_4$ hydrates do not clearly show this expected trend. The lattice parameters of deuterated $CH_4$ hydrates are slightly larger than that of hydrogenated ones and do not show any dependency on the cage filling. The difference in lattice constants (and thus in cage volume) likely at the origin of the higher SC filling in the deuterated hydrates; the 3 K difference in temperature can be negligible.

| No. | Cage type | T K | P MPa | $R_{wp}$ % | a Å | $\theta^D_{LC}$ % | $\theta^D_{SC}$ % | $\theta^D_{LC}/\theta^D_{SC}$ | n | Ice wt.% | $N_d$ |
|---|---|---|---|---|---|---|---|---|---|---|---|
| 1 | $D_2O$ | 271 | 3.5 | 4.80 | 11.85808(2) | 99.6(1.0) | 85.8(1.3) | 1.16(3) | 5.98(6) | 62.76(7) | 5 |
| 2 | $D_2O$ | 271 | 6 | 7.95 | 11.85688(2) | 98.9(3) | 85.6(4) | 1.16(1) | 6.02(2) | 32.86(6) | 9 |
| 3 | $D_2O$ | 271 | 10 | 12.89 | 11.85661(6) | 100.0(5)[a] | 87.3(8) | 1.15(1) | 5.94(4) | 35.63(13) | 3 |
| 4 | $D_2O$ | 271 | 15 | 8.92 | 11.85770(4) | 100.0(5)[a] | 88.5(4) | 1.13(1) | 5.92(2) | 18.77(7) | 3 |
| 5 | $H_2O$ | 268 | 3.5 | 8.28 | 11.85474(2) | 98.3(3) | 85.6(4) | 1.15(1) | 6.04(2) | 30.45(6) | 3 |
| 6 | $H_2O$ | 268 | 6 | 8.90 | 11.85627(2) | 100.0(5)[a] | 87.1(5) | 1.16(1) | 5.94(3) | 38.23(8) | 3 |
| 7 | $H_2O$ | 268 | 10 | 15.29 | 11.85448(11) | 100.0(5)[a] | 87.0(9) | 1.15(2) | 5.94(4) | 46.45(12) | 3 |
| 8 | $H_2O$ | 268 | 15 | 8.53 | 11.85484(4) | 100.0(5)[a] | 85.4(4) | 1.17(1) | 5.97(2) | 27.71(7) | 3 |

Table 1 Crystal structure analysis of sI $CH_4$ hydrates: lattice parameter *a*, the cage filling $\theta$ and hydration number *n*. $R_{wp}$ and $N_d$ are the weighted R-value and number of scans (esd's are quoted in parentheses), respectively. [a] $\theta$ was fixed to 100%, and esd's were estimated.

**Raman spectroscopy**

As shown in Figure 2, when $CH_4$ molecules are encaged into the sI hydrogenated or deuterated hydrates cavities, the Raman band of the totally symmetric stretching-vibration mode of C-H splits into two peaks at ~2901 and 2913 $cm^{-1}$, assigned to $CH_4$ in the LCs and SCs, respectively [6]. A significant background change from $CH_4$ gas to the deuterated $CH_4$ hydrates can be observed in the 2830-3200 $cm^{-1}$ range. It is likely to result from guest-host interactions. A similar behavior is expected for the hydrogenated hydrates, which cannot be observed due to the overlapping with the O-H stretching bands. The large-to-small cage occupancy ratio of $CH_4$ in sI $CH_4$ hydrates can be obtained by,

$$\theta_{LC}/\theta_{SC} = \frac{A_{LC} \times \sigma_{SC}}{3A_{SC} \times \sigma_{LC}} \quad (2a)$$

$$\approx A_{LC}/3A_{SC} \quad (2b)$$

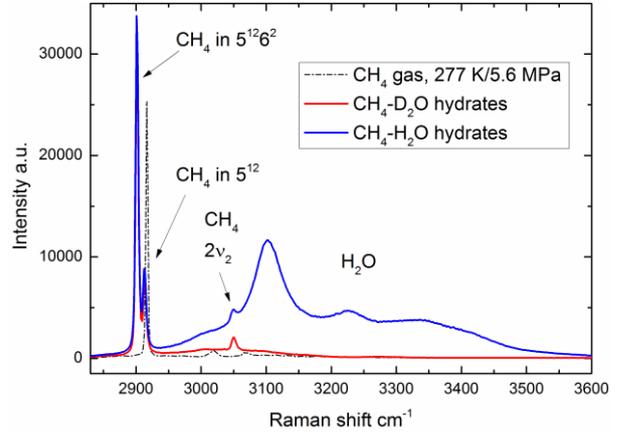

Figure 2 Typical Raman spectra of $CH_4$ hydrates and $CH_4$ gas over the wavenumber of 2830-3600.

where *A* and $\sigma$ are the integrated peak area within the specific range and the corresponding Raman scattering cross section, respectively. The average $\sigma_{LC}/\sigma_{SC}$ for the investigated samples is ~0.97 within 2830-3000 $cm^{-1}$ [6]. In this study, to compare the results determined by synchrotron X-ray diffraction with Raman measurements, only Eqn. 2b is considered. The relative cage occupancies of $CH_4$ in the hydrogenated hydrates are in a good agreement with those of the deuterated hydrates over 2830-3000 $cm^{-1}$, but several percent higher than those over 2830-3600 $cm^{-1}$, see table 2. The lower ratios may be caused by the above-mentioned background change. Therefore, the cage occupancy ratios of $CH_4$-$H_2O$ hydrates derived over 2830-3000 $cm^{-1}$ are thought to be more accurate.

The absolute cage occupancies of $CH_4$ in the LCs and SCs were calculated from the Raman peak intensities, using the relative Raman quantification factors (RQFs) of $CH_4$ to $H_2O/D_2O$ [6,60], or the statistical thermodynamic expression (Eqn. 3). The peak intensities of $CH_4$ and $H_2O$ of $CH_4$-$H_2O$ hydrates integrated over 2830-3600 cm$^{-1}$ and the corresponding RQFs corrected for the presence of ice Ih [60] were used to determine the absolute cage occupancy, while the peak areas of $CH_4$ over 2830-3000 cm$^{-1}$ and $D_2O$ over 2100-2830 cm$^{-1}$ and the relative RQFs were applied for $CH_4$-$D_2O$ hydrates [6]. Due to the inhomogeneous nature of the ice-to-hydrate conversion reaction a sample spot with high intensity of $CH_4$ was manually selected, but the presence of some ice in the focal spot could not be excluded; the straightforward application of RQFs established in [6] is therefore not possible. To solve this problem, the large cage fillings were set to 99%, a value close to the diffraction results (Table 1). With this assumption, the SC fillings, as well as the ice concentration in the measured particles can be determined, see entries in Table 2 with grey background. The estimated percentages of ice in the local spots measured by Raman scattering are much lower than the space-averaged results determined by diffraction (Table 1). This procedure can be applied to *in situ* estimations of the concentration of ice Ih in natural methane hydrate reservoirs, and avoid the overestimation of the hydration number. If water or seawater exists, the difference in the RQF can introduce larger uncertainty [61].

| No. | $\theta_{LC}^{R1}/\theta_{SC}^{R1}$ | RQF | | | $\theta_{LC}^{R2}/\theta_{SC}^{R2}$ | Thermodynamic expression | | | $N_d$ |
| --- | --- | --- | --- | --- | --- | --- | --- | --- | --- |
| | | $\theta_{LC}^{R}$ % | $\theta_{SC}^{R}$ % | Ice wt.% | | $\theta_{LC}^{R}$ % | $\theta_{SC}^{R1}$ % | $n$ | |
| 1 | - | 99 | 86(1) | 10(10) | 1.148(14) | 98.99 | 86.23 | 5.98 | 15 |
| 2 | - | 99 | 87(1) | 6 (6) | 1.125(11) | 98.95 | 87.95 | 5.95 | 15 |
| 3 | - | 99 | 89(3) | 6(10) | 1.100(23) | 98.88 | 89.89 | 5.92 | 15 |
| 4 | - | 99 | 91(2) | 7(8) | 1.077(20) | 98.80 | 91.74 | 5.90 | 16 |
| 5 | 1.09(2) | 99 | 84(1) | 19(8) | 1.150(17) | 98.87 | 85.97 | 5.98 | 16 |
| 6 | 1.07(2) | 99 | 86(2) | 28(15) | 1.135(26) | 98.83 | 87.08 | 5.97 | 12 |
| 7 | 1.05(2) | 99 | 87(1) | 12(5) | 1.110(24) | 98.77 | 88.98 | 5.94 | 14 |
| 8 | 1.04(2) | 99 | 88(2) | 13(6) | 1.102(21) | 98.75 | 89.61 | 5.93 | 11 |

Table 2 Cage occupancies, cage occupancy ratios and hydration number of $CH_4$ hydrates determined by Raman. $\theta_{LC}^{R1}/\theta_{SC}^{R1}$ was calculated by the peak areas integrated over 2830-3600 cm$^{-1}$ for the hydrogenated hydrates. $\theta_{LC}^{R2}/\theta_{SC}^{R2}$ integrated over 2830-3000 cm$^{-1}$ was previously report [6].

In the equilibrated ice-gas-hydrate system, assuming (1) the guest-guest interactions are negligible; (2) the host cages are rigid and one cage only holds one guest molecule; (3) classical statistics are valid [40], the chemical potential difference between the metastable empty hydrate and filled sI hydrate cavities is expressed as,

$$\Delta\mu^H(T,P) = \mu^\beta(T_0,P_0) - \mu^H(T,P)$$
$$= -\frac{RT}{23}\left[3\ln(1-\theta_{LC}) + \ln(1-\theta_{SC})\right] \quad (3)$$

where $\mu^\beta(T_0, P_0)$ is the chemical potential of the hypothetical empty hydrate at reference temperature $T_0$ and pressure $P_0$, usually taken as 273.15 K and 0 MPa, and $\mu^H(T, P)$ denotes the chemical potential of water in the hydrate phase under certain temperature and pressure. Obviously, if the absolute cage occupancies are accurately known for given formation conditions, $\Delta\mu^H$ can be directly derived. With the unprecedented precision of our synchrotron diffraction results, it may appear tempting to calculate $\Delta\mu^H$ using Eqn. 3. However, as the chemical potential varies with the cage filling in a logarithmic function, for fractional fillings approaching 100%, $\Delta\mu^H$ becomes very sensitive to $\theta$'s (Eqn. 3). Assuming the LC occupancies larger than 99% in Table 1 are 99%, the averaged $\Delta\mu^H$'s for the deuterated and hydrogenated hydrates derived from Eqn. 3 are 1546±22 and 1493±81 J/mol, respectively; the quoted standard reflects the deviation from average and does not reflect the much bigger uncertainty of any systematic error. As the true $\mu^\beta(T_0, P_0)$ of the hypothetic sI empty hydrate is not known with certainty, e.g. [62,63], these averaged $\Delta\mu^H$'s were substituted into Eqn. 3, in collaboration with the cage occupancy ratio determined by Raman spectra integrated over 2830-3000 cm$^{-1}$ (Table 2), to tentatively determine the absolute cage occupancy. The results of methane hydrate stoichiometry are close to 6 for both diffraction and Raman data, see Table 1 and 2.

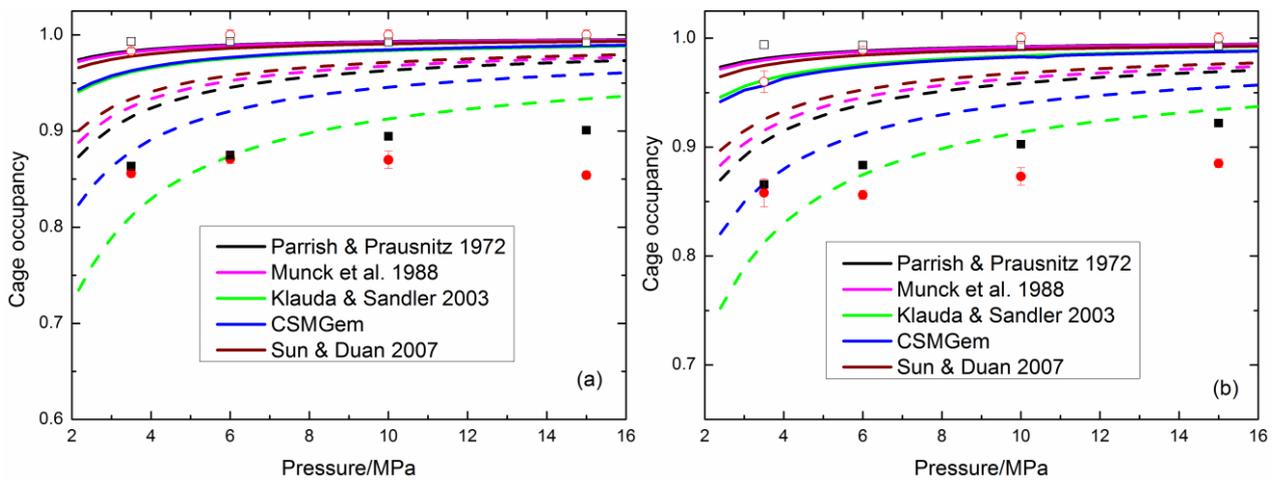

Figure 3 Cage occupancies of $CH_4$ in sI hydrogenated (a) and deuterated $CH_4$ hydrates (b) determined by diffraction patterns (circle) and Raman spectra (square) versus pressure plot, and comparison with the predictions of the thermodynamic models. Empty and solid symbols correspond to the LC and SC, respectively.

**Comparison of absolute cage occupancies**

As it can be seen in Table 2, the absolute cage occupancies determined by the RQFs and statistical thermodynamic expression agree well. The difference in the SC fillings determined by Raman and diffraction vary from being almost identical at 3.5 MPa up to 5% at 15 MPa, see Figure 3. This difference likely originates in the missing correction for the Raman cross sections of $CH_4$ in the LCs to SCs [6], or (less likely) by the remaining systematic errors in the Rietveld analysis arising from the parameter correlations.

As shown in Figure 3, the LCs are nearly completely filled for both the experimental determinations and the predictions. However, thermodynamic models seem to be inadequate to predict the SC filling, as it can be seen from the significant discrepancies among models. The data show a clear non-Langmuir behavior for the SC filling and a filling which is substantially lower than all predictions at higher pressure. The disagreement between the thermodynamic models is striking and a result of differently approximated guest-host/guest interactions. Given reliable experimental data, the capability to reproduce the experimental cage occupancy is a very critical test for a given prediction model. For methane hydrates formed at 3.5 and 6 MPa, our diffraction and Raman results agree reasonably well with CSMGem and Klauda & Sandler [47], and are inconsistent with all other models. With increasing formation pressure, the diffraction results become up to 9% lower than Klauda & Sandler [47] which predicts the lowest populations of $CH_4$ in the SCs and up to 11% lower than CSMGem, while the Raman results still remain comparable. These large discrepancies in the SC-filling between models and our results can be caused by all or one of the following three aspects: the inadequacy of the prediction model(s), inaccuracies of the experimental approach and/or an incomplete equilibration during the formation of the hydrates investigated. While a clear answer cannot be given at this stage, some further comments may be given. Assuming that our ice-gas-hydrate system does not reach equilibrium, the lower SC-fillings can be plausibly interpreted by the permeation limitation of guest molecules, which could lead to an increasingly lower cage filling despite the higher driving forces at higher pressure [64]. Two effects can contribute to this phenomenon. With increasing degree of transformation at higher pressures the distance to the reaction front increases which slows down the equilibration process. Moreover, $CH_4$ hydrates with higher cage occupancies were initially formed at the outer layer of ice particle in the ice-gas system at higher pressure. As empty cages are important for the diffusion of guest molecules through the hydrate lattice [64], this interface with only a few empty cages hinders the in-diffusion of $CH_4$ molecules towards the ice core. Considering the very low diffusion constant of $CH_4$ in hydrates, in the order of $\sim 10^{-15}$ m$^2$/s or lower [64,65], the lower small cage occupancy determined by high-resolution diffraction may be due to the un-reached phase equilibrium. The case of $CO_2$ hydrates lend further support to this suggestion – the absolute cage occupancies of $CO_2$ hydrates formed in the ice-gas system at 1.5 and 3 MPa are consistent with the predictions by CSMGem [20]; due to the ~3 times higher diffusion constant of $CO_2$ in a hydrate lattice the equilibration process will be faster [64,67].

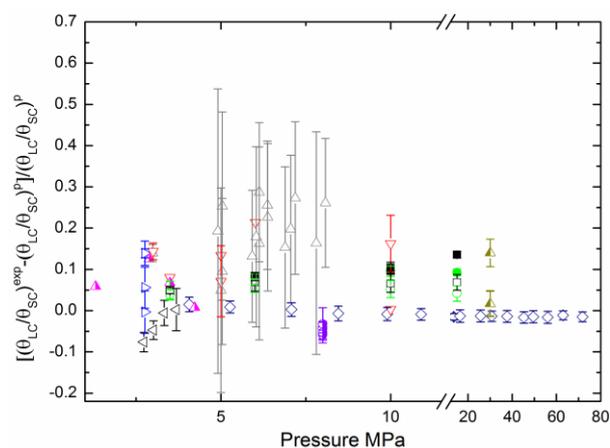

Figure 4 Pressure plot versus difference in the cage occupancy ratio of $CH_4$ in the synthetic sI $CH_4$ hydrates and the calculated values by CSMGem under the same *T-P*. ◁ [8], △ [7], ▷ [29], ◇ [22], ▲ [38], ▲ [21], ◐ [24], ▽ [23], [28], [39], [26], [27], [25] (left to right). Circle and square denote $CH_4$-$H_2O$ and $CH_4$-$D_2O$ hydrates; solid and empty are the diffraction (this work) and Raman [6] results, respectively. The superscripts "exp" and "p" are the experimental and calculated values, respectively.

**Comparison of cage occupancy ratios**
To give a better view of the difference between experimental values and predictions in the cage occupancy ratio, the results for both the experimentally-determined values of sI synthetic $CH_4$ hydrates (Figure 4) as well as the sI natural methane hydrates (Figure 5) are compared with the calculations from CSMGem [44], which has successfully predicted $\theta_{LC}/\theta_{SC}$ [55]; CSMGem well reproduces some experimental results in references [8,21,22,38], particularly Jager's work in which the $\theta_{LC}/\theta_{SC}$'s of $CH_4$ hydrates formed from the water-gas system under pressures close to the dissociation pressures were determined by Raman spectra [22]. However, the capability to reproduce other results [7,8,21,23-29,36,39] including the results from our study is relatively poor. It is still difficult to draw final conclusions on the actual cage occupancy ratios of methane hydrates due to the scattered experimental formation conditions, the unknown quality of measurements and ambiguities in the way the data are analyzed. As the formation of $CH_4$ hydrates in the ice-gas system is a very slow, permeation-controlled process, the reaction time is determining whether the formation starting from ice has reached equilibrium or not. The cage occupancy ratio of methane hydrates synthesized from the ice-gas system for a few weeks were found ~21% lower than the prediction by CSMGem [26]. Subramanian's results show that the cage occupancy ratio of $CH_4$ hydrates formed for 2 years are ~13% lower than the one for a sample reacted for two months using NMR [29]. Furthermore, this $\theta_{LC}/\theta_{SC}$ value is identical with the prediction by CSMGem, see Figure 4. Based on our above-mentioned discussion and these observations, we come to the preliminary conclusion that $CH_4$ hydrates formed in the ice-gas system for several days to weeks may not have reached equilibrium. Consequently, when deriving the absolute cage occupancy from Eqn. 3 and the cage occupancy ratio, the wrongly assumed equilibration may lead to an overestimation of the actual SC occupancy.

The cage occupancy ratio of $CH_4$ hydrates synthesized from the water-gas system was found to be independent of the reaction time [21], which suggests the water-gas-hydrate equilibrium can be reached shortly and the formation mechanism is not a permeation-controlled process. Moreover, the cage occupancy ratio of $CH_4$ hydrates formed at the interface of water and vapor phase is ~11% lower than the dendrite hydrate formed underneath the interfacial hydrates and in the water-rich phase [21]. All the predictions by CSMGem in this study were calculated from the water/ice-gas system with the composition of 90 mol% $CH_4$ and 10 mol% $H_2O$, based on the following concerns: 1) it is difficult to trace back to the measured positions whether in the gas-water interfacial layer or in the water-rich phase in the published literatures; 2) the gas-rich phase also locally exists as gas conduit has been observed in NGH veins [66]; 3) under some *x-P-T* conditions, the cage occupancy cannot be calculated by CSMGem. The available $\theta_{LC}/\theta_{SC}$'s are usually higher than the CSMGem predictions, see Figure 4. Clearly, there are major uncertainties what concerns the expected range of $\theta_{LC}/\theta_{SC}$ ratios as a function of the formation process (water or gas excess). Moreover, it is not clear by which process natural gas hydrates are formed, nor even whether the hydrate formation from ice can be considered as a process with excess gas.

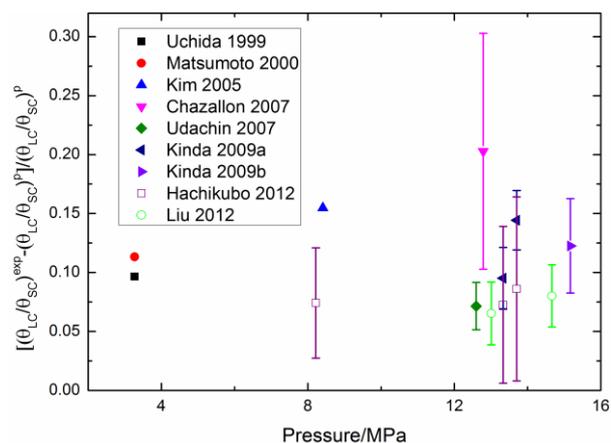

Figure 5 Pressure plot versus difference in the cage occupancy ratios of $CH_4$ in natural $CH_4$-hydrates and the calculated values by CSMGem under the same *T-P*. In [30], water depth is set as 1200 m.

Naturally occurring methane hydrates in marine environments are certainly be formed in the water-gas system (either locally gas- or water-rich) and can be expected to have reached equilibrium. In this study, only guest compositions with more than 99% $CH_4$ and no reported $H_2S$ and $N_2$ which are known to significantly influence the cage populations of $CH_4$ [44] are considered. As shown in Figure 5, all available $\theta_{LC}/\theta_{SC}$ data of NMHs are higher than the predictions by CSMGem. The *P-T* data in the gas hydrate reservoirs were directly taken from the references, or estimated from the sea water depth and sample depth below the sea-floor; it should be noted that in some cases accurate *P-T* data of the retrieved samples were not available. There is also a general uncertainty about the formation conditions, in particular the presence or absence of free gas, for a number of MGH; these conditions will also affect the LC/ SC ratios. Nevertheless, the fact that for the natural $CH_4$ hydrates (which are likely to be better equilibrated) the observed ratios are generally larger than the predictions suggests a deficiency in the prediction models.

**CONCLUSIONS**

The compositional and structural information of sI $CH_4$-$H_2O$ and $CH_4$-$D_2O$ hydrates determined by high-resolution synchrotron diffraction and Raman spectroscopy were reported. The LCs are found to be almost fully occupied, which agrees with the thermodynamic models. Our diffraction results show the SC fillings in the deuterated hydrates are slightly higher than that in the hydrogenated ones. This is ascribed to the slightly larger lattice parameters in the deuterated form. The relative Raman RQFs of $CH_4$ to $H_2O$/$D_2O$ or statistical thermodynamic expression, in cooperation with Raman intensities, are both used to determine the absolute cage occupancies. The Raman results agree well with each other, and are consistent with the diffraction results.

There are large discrepancies in the predicted cage fillings among the various thermodynamic models; only CSMGem and the *ab-initio* based model are close to the experimental evidence at low driving force. However, the non-Langmuir behavior of the SC-occupancies with increasing $CH_4$ formation fugacity may well be due to the lack of equilibration of the $CH_4$ fillings during the preparation period of 3 weeks, in particular for the SCs. Both, the published and here determined LC/SC occupancy ratios of the synthetic and natural $CH_4$ hydrates were generally found to be higher than the values predicted. The reason for these discrepancies remain unresolved; part of the uncertainty is due to the fact that the formation conditions apparently influence this ratio [21] combined with the fact that the exact formation conditions are of unknown, in particular for natural hydrates. Further NMR or *ex situ* Raman measurements are suggested to investigate the cage-occupancy difference between methane hydrates formed in the gas-water interfacial and in the water-rich phase. Furthermore, the deconvolution of *in situ* acquired Raman spectra with contributions of free gas, gas-saturated water-rich phase and gas hydrates introduces additional uncertainties. *Ex situ* Raman and diffraction analyses of $CH_4$-hydrate sample formed in the water-rich phase of the water-gas-hydrate-sand system and recovered from NMH reservoirs with no detectable gas conduits are under way.

**REFERENCES**

[1] Kvenvolden KA. *Potential effects of gas hydrate on human welfare*. Proc. Natl. Acad. Sci. U.S.A. 1999;96(7):3420-3426.

[2] Ohgaki K, Takano K, Sangawa H, Matsubara T, Nakano S. *Methane exploitation by carbon dioxide from gas hydrates - Phase equilibria for $CO_2$-$CH_4$ mixed hydrate system*. J. Chem. Eng. Jpn. 1996;29(3):478-483.

[3] Dickens GR. *Rethinking the global carbon cycle with a large, dynamic and microbially mediated gas hydrate capacitor*. Earth and Planetary Science Letters. 2003;213(3-4):169-183.

[4] Deaton WM, Frost EMJ. *Gas hydrates and their relation to the operation of natural-gas pipe lines*: United States Department of the Interior, Bureau of Mines, 1946.

[5] Sloan ED, Koh C, A. *Clathrate hydrates of natural gases*. 3rd ed. Boca Raton, FL: CRC Press, Taylor & Francis Group, 2008.

[6] Qin J, Kuhs WF. *Quantitative analysis of gas hydrates using Raman spectroscopy*. AIChE J. 2013;59(6):2155-2167.

[7] Uchida T, Hirano T, Ebinuma T, Narita H, Gohara K, Mae S, Matsumoto R. *Raman spectroscopic determination of hydration number of methane hydrates*. AlChE J. 1999;45(12):2641-2645.

[8] Sum AK, Burruss RC, Sloan ED. *Measurement of clathrate hydrates via Raman spectroscopy*. J. Phys. Chem. B. 1997;101(38):7371-7377.

[9] Glew DN. *Aqueous solubility and gas-hydrates - methane-water system*. J. Phys. Chem. 1962;66(4):605-609.

[10] de Roo JL, Peters CJ, Lichtenthaler RN, Diepen GAM. *Occurrence of methane hydrate in saturated and unsaturated solutions of sodium-chloride and water in dependence of temperature and pressure*. Aiche Journal. 1983;29(4):651-657.

[11] Handa YP. *Compositions, enthalpies of dissociation, and heat-capacities in the range 85 k to 270 k for clathrate hydrates of methane, ethane, and propane, and enthalpy of dissociation of isobutane hydrate, as determined by a heat-flow calorimeter*. J. Chem. Thermodyn. 1986;18(10):915-921.

[12] Circone S, Kirby SH, Stern LA. *Direct measurement of methane hydrate composition along the hydrate equilibrium boundary*. J. Phys. Chem. B. 2005;109(19):9468-9475.

[13] Galloway TJ, Ruska W, S. CP, Kobayash R. *Experimental measurement of hydrate numbers for methane and ethane and comparison with theoretical values*. Ind. Eng. Chem. Fundam. 1970;9(2):237-243.

[14] Takeya S, Kida M, Minami H, Sakagami H, Hachikubo A, Takahashi N, Shoji H, Soloviev V, Wallmann K, Biebow N, Obzhirov A, Salomatin A, Poort J. *Structure and thermal expansion of natural gas clathrate hydrates*. Chem. Eng. Sci. 2006;61(8):2670-2674.

[15] Milkov AV. *Global estimates of hydrate-bound gas in marine sediments: how much is really out there?* Earth Sci. Rev. 2004;66(3-4):183-197.

[16] Koh CA, Sloan ED, Sum AK, Wu DT. *Fundamentals and applications of gas hydrates*. Annu Rev Chem Biomol Eng. 2011;2:237-257.

[17] Takeya S, Udachin KA, Moudrakovski IL, Susilo R, Ripmeester JA. *Direct space methods for powder X-ray diffraction for guest-host materials: Applications to cage occupancies and guest distributions in clathrate hydrates*. J. Am. Chem. Soc. 2010;132(2):524-531.

[18] Udachin KA, Lu H, Enright GD, Ratcliffe CI, Ripmeester JA, Chapman NR, Riedel M, Spence G. *Single crystals of naturally occurring gas hydrates: The structures of methane and mixed hydrocarbon hydrates*. Angew. Chem. Int. Ed. 2007;46(43):8220-8222.

[19] Udachin KA, Ratcliffe CI, Ripmeester JA. *Structure, composition, and thermal expansion of $CO_2$ hydrate from single crystal X-ray diffraction measurements*. J. Phys. Chem. B. 2001;105(19):4200-4204.

[20] Hartmann CD, SHemes S, Falenty A, Kuhs WF. *The structure and cage filling of gas hydrates as established by synchrotron powder diffraction data*. In: Proceedings of the 7th International Conference on Gas Hydrates, Edinburgh, Scotland, United Kingdom, July 17-21, 2011.

[21] Huo ZX, Hester K, Sloan ED, Miller KT. *Methane hydrate nonstoichiometry and phase diagram*. AIChE J. 2003;49(5):1300-1306.

[22] Jager MD. *High pressure studies of hydrate phase inhibition using Raman spectroscopy*, Colorado School of Mines; 2001.

[23] Chazallon B, Focsa C, Charlou JL, Bourry C, Donval JP. *A comparative Raman spectroscopic study of natural gas hydrates collected at different geological sites*. Chem. Geol. 2007;244(1-2):175-185.

[24] Liu C, Lu H, Ye Y, Ripmeester JA, Zhang X. *Raman spectroscopic observations on the structural characteristics and dissociation behavior of methane hydrate synthesized in silica sands with various sizes*. Energy & Fuels. 2008;22(6):3986-3988.

[25] Liu C, Ye Y, Meng Q, He X, Lu H, Zhang J, Liu J, Yang S. *The characteristics of gas hydrates

*recovered from shenhu area in the South China Sea*. Mar. Geol. 2012;307:22-27.
[26] Ohno H, Kida M, Sakurai T, Iizuka Y, Hondoh T, Narita H, Nagao J. *Symmetric stretching vibration of $CH_4$ in clathrate hydrate structures*. ChemPhysChem. 2010;11(14):3070-3073.
[27] Seo Y, Lee H, Ryu BJ. *Hydration number and two-phase equilibria of $CH_4$ hydrate in the deep ocean sediments*. Geophys. Res. Lett. 2002;29(8).
[28] Uchida T, Takeya S, Ebinuma T, Narita H. *Replacing methane with $CO_2$ in clathrate hydrate: observations using Raman spectroscopy. In: Proceedings of the 5th International Conference on Gas Hydrates, Cairns, Australia, 2001*.
[29] Subramanian S. *Measurements of clathrate hydrates containing methane and ethane using Raman spectroscopy*, Colorado School of Mines; 2000.
[30] Kida M, Suzuki K, Kawamura T, Oyama H, Nagao J, Ebinuma T, Narita H, Suzuki H, Sakagami H, Takahashi N. *Characteristics of Natural Gas Hydrates Occurring in Pore-Spaces of Marine Sediments Collected from the Eastern Nankai Trough, off Japan*. Energy & Fuels. 2009;23:5580-5586.
[31] Hachikubo A, Khlystov O, Kida M, Sakagami H, Minami H, Yamashita S, Takahashi N, Shoji H, Kalmychkov G, Poort J. *Raman spectroscopic and calorimetric observations on natural gas hydrates with cubic structures I and II obtained from Lake Baikal*. Geo-Mar. Lett. 2012;32(5-6):419-426.
[32] Kim DY, Uhm TW, Lee H, Lee YJ, Ryu BJ, Kim JH. *Compositional and structural identification of natural gas hydrates collected at site 1249 on ocean drilling program leg 204*. Korean J. Chem. Eng. 2005;22(4):569-572.
[33] Matsumoto R, Uchida T, Waseda A, Uchida T, Takeya S, Hirano T, Yamada K, Maeda Y, Okui T. *2. Occurrence, structure, and composition of natural gas hydrate recovered from the blake ridge, northwest atlantic. In: Proc. ODP, Sci. Results, College Station, TX, 2000*.
[34] Wilson LD, Tulk CA, Ripmeester JA. *Instrumental techniques for the investigation of methane hydrate: cross-callibrating NMR and Raman spectroscopic data. In: Proceedings of the 4th International Conference on Gas Hydrates, Yokohama, Japan, May 19-23, 2002*.
[35] Uchida T, Takeya S, Wilson LD, Tulk CA, Ripmeester JA, Nagao J, Ebinuma T, Narita H. *Measurements of physical properties of gas hydrates and in situ observations of formation and decomposition processes via Raman spectroscopy and X-ray diffraction*. Can. J. Phys. 2003;81(1-2):351-357.
[36] Gupta A, Dec SF, Koh CA, Sloan ED. *NMR investigation of methane hydrate dissociation*. J. Phys. Chem. C. 2007;111(5):2341-2346.
[37] Kida M, Hachikubo A, Sakagami H, Minami H, Krylov A, Yamashita S, Takahashi N, Shoji H, Khlystov O, Poort J, Narita H. *Natural gas hydrates with locally different cage occupancies and hydration numbers in Lake Baikal*. Geochemistry Geophysics Geosystems. 2009;10.
[38] Kini RA. *NMR studies of $CH_4$, $C_2H_6$, and $C_3H_8$ hydrates: structure, kinetics, and thermodynamics* Colorado School of Mines; 2002.
[39] Seo YT, Lee H. *C-13 NMR analysis and gas uptake measurements of pure and mixed gas hydrates: Development of natural gas transport and storage method using gas hydrate*. Korean J. Chem. Eng. 2003;20(6):1085-1091.
[40] van der Waals JH, Platteeuw JC. *Clathrate solutions*. Adv. Chem. Phys. 1959;2:1-57.
[41] Parrish WR, Prausnit JM. *Dissociation pressures of gas hydrates formed by gas-mixtures*. Ind. Eng. Chem. Process Des. Dev. 1972;11(1):26-35.
[42] Munck J, Skjoldjorgensen S, Rasmussen P. *Computations of the formation of gas hydrates*. Chem. Eng. Sci. 1988;43(10):2661-2672.
[43] Ballard AL, Sloan ED. *The next generation of hydrate prediction I. Hydrate standard states and incorporation of spectroscopy*. Fluid Phase Equilibria. 2002;194:371-383.
[44] Ballard AL, Sloan ED. *The next generation of hydrate prediction - Part III. Gibbs energy minimization formalism*. Fluid Phase Equilibria. 2004;218(1):15-31.
[45] Cao ZT, Tester JW, Sparks KA, Trout BL. *Molecular computations using robust hydrocarbon-water potentials for predicting gas hydrate phase equilibria*. J. Phys. Chem. B. 2001;105(44):10950-10960.
[46] Klauda JB, Sandler SI. *Ab initio intermolecular potentials for gas hydrates and their predictions*. J. Phys. Chem. B. 2002;106(22):5722-5732.
[47] Klauda JB, Sandler SI. *Phase behavior of clathrate hydrates: a model for single and multiple gas component hydrates*. Chem. Eng. Sci. 2003;58(1):27-41.
[48] Sun R, Duan ZH. *Prediction of $CH_4$ and $CO_2$ hydrate phase equilibrium and cage occupancy*

from ab initio intermolecular potentials. Geochim. Cosmochim. Acta. 2005;69(18):4411-4424.

[49] Sun R, Duan Z. *An accurate model to predict the thermodynamic stability of methane hydrate and methane solubility in marine environments*. Chem. Geol. 2007;244(1-2):248-262.

[50] Anderson BJ, Tester JW, Trout BL. *Accurate potentials for argon-water and methane-water interactions via ab initio methods and their application to clathrate hydrates*. J. Phys. Chem. B. 2004;108(48):18705-18715.

[51] Thomas C, Picaud S, Ballenegger V, Mousis O. *Sensitivity of predicted gas hydrate occupancies on treatment of intermolecular interactions*. J. Chem. Phys. 2010;132(10).

[52] Wierzchowski SJ, Monson PA. *Calculation of free energies and chemical potentials for gas hydrates using Monte Carlo simulations*. J. Phys. Chem. B. 2007;111(25):7274-7282.

[53] Sizov VV, Piotrovskaya EM. *Computer simulation of methane hydrate cage occupancy*. J. Phys. Chem. B. 2007;111(11):2886-2890.

[54] Jensen L, Thomsen K, von Solms N, Wierzchowski S, Walsh MR, Koh CA, Sloan ED, Wu DT, Sum AK. *Calculation of liquid water-hydrate-methane vapor phase equilibria from molecular simulations*. J. Phys. Chem. B. 2010;114(17):5775-5782.

[55] Ballard L, Sloan ED. *The next generation of hydrate prediction IV - A comparison of available hydrate prediction programs*. Fluid Phase Equilibria. 2004;216(2):257-270.

[56] Staykova DK, Kuhs WF, Salamatin AN, Hansen T. *Formation of porous gas hydrates from ice powders: diffraction experiments and multistage model*. J. Phys. Chem. B. 2003;107(37):10299-10311.

[57] Larson AC, Von Dreele RB. *General structure analysis system (GSAS)*: Los Alamos National Laboratory, 2004.

[58] Klapproth A. *Strukturuntersuchungen an methanund kohlenstoffdioxid-clathrat-hydraten*, Georg-August-Universität Göttingen; 2002.

[59] Hamilton WC. *Significance tests on crystallographic r factor*. Acta Crystallogr. 1965;18:502-510.

[60] Qin J, Kuhs WF. *Recovering $CH_4$ from clathrate hydates with $CO_2 + N_2$ gas mixtures: Quantitative Raman study. In: Proceedings of the 8th International Conference on Gas Hydrates, Beijing, China, 2014*.

[61] Slusher RB, Derr VE. *Temperature-dependence and cross-sections of some stokes and anti-stokes Raman lines in ice Ih*. Applied Optics. 1975;14(9):2116-2120.

[62] Davidson DW, Handa YP, Ripmeester JA. *Xe-129 NMR and the thermodynamic parameters of Xenon hydrate*. J. Phys. Chem. 1986;90(24):6549-6552.

[63] Lee SY, Holder GD. A generalized model for calculating equilibrium states of gas hydrates: Part II. In: Holder GD, Bishnoi PR, eds. *Gas Hydrates: Challenges for the Future.* Vol 9122000:614-622.

[64] Falenty A, Hansen TC, Salamatin AN, Kuhs WF. *Mechanism of gas migration in clathrate hydrates. In: Proceedings of the 8th International Conference on Gas Hydrates, Beijing, China, 2014.*

[65] Peters B, Zimmermann NER, Beckham GT, Tester JW, Trout BL. *Path sampling calculation of methane diffusivity in natural gas hydrates from a water-vacancy assisted mechanism*. J. Am. Chem. Soc. 2008;130(51):17342-17350.

[66] Abegg F, Bohrmann G, Freitag J, Kuhs W. *Fabric of gas hydrate in sediments from hydrate ridge-results from ODP Leg 204 samples*. Geo-Mar. Lett. 2007;27(2-4):269-277.

**ACKNOWLEDGEMENTS**

The authors thank the European Synchrotron Radiation Facility (ESRF) in Grenoble for beam-time and support, Heiner Bartels and Eberhard Hensel (Göttingen) for technical support, Dr. Andrzej Falenty and Susanne Hemes for sample preparation and performing the experiment on ID31, and Michela Brunelli (ESRF) for help and advice on ID31. This research was supported by the German Ministry of Education and Research (BMBF) in the framework of the special program GEOTECHNOLOGIEN and within the SUGAR-II research initiative by the grant 03G0819B (TP B2-3).